\documentclass[aps,prl,twocolumn,floatfix]{revtex4}

\usepackage{bm,bbm,amsmath, amssymb,amsbsy, amsthm,graphicx,subfigure,color,txfonts,multirow,makecell,relsize,enumitem}
\usepackage[framemethod=tikz]{mdframed}
\usepackage[colorlinks]{hyperref}
\hypersetup{
	bookmarksnumbered,
	pdfstartview={FitH},
	citecolor={blue},
	linkcolor={red},
	urlcolor={black},
	pdfpagemode={UseOutlines}
}
\usepackage{cleveref}

\newcommand{\ket}[1]{| #1 \rangle}
\newcommand{\bra}[1]{\langle #1 |}

\begin{document}
 \title{Quantifying the difference between many-body quantum states}
 \author{Davide Girolami}
\email{davegirolami@gmail.com}
 \affiliation{$\hbox{DISAT, Politecnico di Torino, Corso Duca degli Abruzzi 24, Torino 10129, Italy}$
 }
\author{Fabio Anz\`a}
\email{fanza@ucdavis.edu}
\affiliation{$\hbox{Complexity Sciences Center, University of California at Davis, One Shields Avenue, Davis (CA) 95616, USA}$
 }

\begin{abstract}

The quantum state overlap is the textbook measure of  the difference between two quantum states. Yet, it  is inadequate to compare the complex configurations of  many-body systems.  The problem is inherited by the widely employed  quantum state fidelity and  related distances.
We introduce the weighted distances,  a new class of information-theoretic measures that  overcome these limitations. They quantify how hard it is to discriminate between two quantum states of many particles, factoring in the structure of the required measurement apparatus.  Therefore, they can be used to evaluate both the theoretical and the experimental performances of complex quantum devices. 
We also show that the newly defined ``weighted Bures length'' between the input and output states of a quantum process  is a lower bound to the experimental cost of the transformation. The result uncovers an exact quantum limit to our ability to convert physical resources into computational ones. %, e.g. entanglement generation schemes.The latter is defined as a function of the required energy, time and number of operations we need to implement a satisfactory discrete step approximation of the process.  

%Evaluating the length of the path between the initial and final states of a quantum algorithm lower bounds the physical resources which are needed to implement the transformation.   %That is, the difference between two quantum states is upper bounded by . %Specifically, the Bures weighted distance between two states lower bounds the size of the shortest algorithm to transform a state into the other.  
 %, e.g.   quantum computers.% and many-body entanglement generation schemes in quantum networks. 
\end{abstract}
%\pacs{3.65}

\date{\today}
 
\maketitle
\noindent{\it Introduction --} Quantum particles are the building blocks of light and matter, but they can display very   complex configurations. An important goal of quantum theory is to describe their differences with simple metrics. % is challenging, due to their many degrees of freedom. % It is then compelling to  express  what we know about them in computable metrics. 
 The state overlap $|\bra{i}j\rangle|$ is the standard proxy to compare two wave functions $\ket{i},\ket j$, and it has a compelling statistical meaning: it quantifies how hard it is to discriminate two pure states via a single quantum measurement \cite{wootters}. 
 The overlap is instrumental to build the Fubini-Study distance $\cos^{-1}|\bra{i}j\rangle|$ \cite{fubini,study}, which evaluates the distinguishability of two quantum states in terms of how far they are in the system Hilbert space.\\

\noindent Unfortunately, the state overlap  is not fully adequate to compare many-body wave functions. %, as it does not take into account their internal structure. 
Very similar states can be flagged as maximally different. For example, there is zero overlap between the  $N-$qubit states $\ket{0}^{\otimes N}, \ket{0}^{\otimes N-1}\ket{1}$, for arbitrarily large $N$.
Moreover, geometrically close states can have very different properties. Transforming  $\ket{0}^{\otimes N}$ into the  entangled ``GHZ'' state $a\ket{0}^{\otimes N}+b\ket{1}^{\otimes N}, |a|,|b|\neq 0,1,$ takes experimental resources that grow with the system size \cite{ghz}, e.g. $O(N)$ operations in gate-based quantum  computers \cite{nielsen},  however big their overlap $|a|$ may be.     
 
\noindent The same issues plague the generalizations of the state overlap that  quantify the difference between  two mixed states $\rho,\sigma$, e.g. the  quantum fidelity $F(\rho,\sigma)=\text{Tr}\,\big|\rho^{1/2}\sigma^{1/2}\big|_1$ \cite{fidelity1,fidelity2}, and related distances \cite{geo1}. This fact is troublesome. % It quantifies the similarity between mixed states in terms of the ability to distinguish them via measurements.
 %. 
As we expect to steadily upsize  quantum technologies, we need trustworthy tools to evaluate the performances of large noisy quantum machines \cite{preskill}. Reconstructing the fidelity between, say,  the target and the  output states of a computation, is often the only  way to certify that a device is truly quantum without accessing its inner workings \cite{fidelity3,fido1,fido2,fido3}. \\%Yet, this index fails to address the complex structure of many-body quantum states.\\

\noindent In this work, we introduce  the weighted distances, a  class of measures for comparing many particle states. A standard,  overlap-based distance quantifies the ability to discriminate two states of a system via a single optimal measurement. %Yet, independently monitoring different partitions of a system provides more information about many particle states. 
Here, we  consider a more general scenario. Cooperating observers  independently monitor different subsystems, evaluating the difference between two preparations of the assigned subsystem by a standard distance.
We construct a weighted sum of these distances, such that the importance of each observer contribution is {\it inversely} proportional to the size of the assigned subsystem. Since the difficulty of performing   measurements is arguably related to the size of the required apparatuses, these quantities weight each contribution  in terms of how easy is it to experimentally implement the related measurement.  
We define a weighted distance as the maximum over all this kind of weighted sums. 
  %if a large measurement apparatus is needed to discriminate two states, they are very similar.%This counterintuitive choice is justified, since  the larger is the most informative measurement apparatus, the more similar are the two states, because it is more difficult to experimentally discriminate them.
The weighted distances satisfy a set of desirable mathematical properties, certifying that they are robust information measures.  %In particular, they  upper bound geometric measures of many-body coherence and quantum correlations, as expected. 
We perform explicit calculations of  interesting case studies, showing that the newly defined weighted Bures length is more informative than the related standard Bures length \cite{bures,uhl}. For example, if a large measurement apparatus is needed to discriminate between two states, their weighted distance is short, because it is experimentally difficult to distinguish one state from the other.  %, while it is no more difficult to compute.  \\%When comparing many-body states, it consistently scales with the number of particles.  
%By construction,  the weighted distances are no more difficult to calculate than standard distances. \\%can be estimated with the very same  statistical methods employed for the standard distances, and lower .   %Hence, it is way easier to calculate the weighted distance of an output state to a target entangled state than quantifying its entanglement.

\noindent Then,  we show that the  weighted Bures length between the input and output states of a  quantum process is a  lower bound to the physical resources that are needed  to implement the transformation. That is, the ability to discriminate two quantum states is never greater than the experimental cost of  transforming one state into the other. The result is surprising: state distinguishability and state transformation are considered ``quite different'' tasks \cite{woot2}. We demonstrate that they are related. Previous works established the minimum time and energy-time (``action'') to perform state transformations \cite{unc,speed,luo,gibi,speed2,speed3}.  The input/output  weighted Bures length is a lower bound  to a newly defined index, which factors the required energy, time and size of gates for quantum state preparation.  While proving the optimality of quantum algorithms is notoriously hard \cite{monta}, the result highlights a fundamental quantitative limit to   quantum information processing.  The bound is also valid for  mixed states and non-unitary state transformations. Hence,  it applies to realistic, noisy quantum dynamics. \\
%A metrological framework builds on distinguishability measures. A complexity framework instead evaluate quantum distinguishability in terms of the number of gates required to transforms two states into one another. The two views seem incompatible.  

\noindent {\it Definition and justification of weighted distances --}  Let us call $\rho_N, \sigma_N$ two arbitrary density matrices that represent different preparations of an $N$ particle quantum system. It is well-known that full reconstruction of  quantum states is a daunting task \cite{tomo}. It is therefore interesting to build an information measure that captures the difficulty to discriminate between the two states with a single measurement.  Suppose one  can  perform all possible POVM (positive operator-valued measure) on  the system: ${\cal M}=\left\{{\cal M}_i\geq0, \sum_i {\cal M}_i=I_N \right\}$ \cite{nielsen}. %\otimes_{i=1}^N{\cal M}_{i,\alpha_i}, \sum_\alpha {\cal M}_{i,\alpha_i}=I, {\cal M}_{i,\alpha_i}\geq 0$. 
 The  ability to distinguish between $\rho_N$ and $\sigma_N$ is customarily quantified via maximization of a certain classical statistical distance $d_{cl}$ for probability distributions \cite{geo1}: 
 \begin{align}\label{eq1}
 d(\rho_N,\sigma_N)&:=\max_{{\cal M}}\sum_i d_{cl}\left(\text{Tr}\left\{{\cal M}_i \rho_N\right\}, \text{Tr}\left\{{\cal M}_i\sigma_N\right\}\right)\nonumber\\
 &:=\sum_i d_{cl}\left(\text{Tr}\left\{\tilde{{\cal M}}_i \rho_N\right\}, \text{Tr}\left\{\tilde{{\cal M}}_i\sigma_N\right\}\right),
 \end{align}  %Important examples are the trace distance and the Bures distance. 
in which $\tilde{{\cal M}}=\left\{\tilde{{\cal M}}_i\right\}$ is the most informative measurement. 
Given three arbitrary density matrices $\rho_N,\sigma_N$ and $\tau_N$, we assume that the quantity  meets the following criteria:
\begin{align}\label{prop}
&d(\rho_N,\sigma_N) \geq 0\,\, \text{(non-negativity)},\\
&d(\rho_N,\sigma_N)=0\iff \rho_N=\sigma_N\,\,\ \text{(faithfulness)},\nonumber\\
&d(\rho_N,\sigma_N)\geq d(\Lambda (\rho_N),\Lambda (\sigma_N))\,\forall\,\Lambda\,\,\text{(contractivity),}\nonumber\\
&d(\rho_N,\sigma_N)\leq  d(\rho_N,\tau_N)+d(\tau_N, \sigma_N)\,\,\text{(triangle inequality)}, \nonumber 
\end{align}
 in which $\Lambda$ is a completely positive trace-preserving (CPTP) map, the most general kind of quantum operation \cite{nielsen}. The distance is normalized  such that it takes the maximal value $M_d$  for orthogonal   states, $d(\rho_N,\sigma_N)=M_d\Leftrightarrow \text{Tr}\{\rho_N \sigma_N\}=0$. Indeed, these states can be discriminated with certainty. Contractivity under CPTP maps implies that the distance is non-increasing under partial trace,  $d(\rho_N,\sigma_N)\geq d(\rho_{k},\sigma_{k})$, in which  $\rho_k,\sigma_k$ are the states of a $k<N$-particle subset. The ability to extract information from quantum systems depends on the size of the measurement setup.  However, the distance function is not explicitly dependent on the number of particles $N$, nor the size of the optimal measurement apparatus $\tilde{{\cal M}} $. Indeed, there are in general several  solutions of the maximization in \cref{eq1}. This degeneracy is maximal for  pairs like the $N$ qubit states $\ket{0}^{\otimes N}, \ket{1}^{\otimes N}$: they are perfectly discriminated by projecting on the computational bases $\{0,1\}^{\otimes k},\,\forall\, k\in[1,N]$. \\%Also, a geometric distance quantifies how different are two states when comparing the probability distributions related to the specific optimal measurement. A more informative metric should evaluate the distinguishability via all possible measurements. %That is, the fact that the observer employs a full scale $N$ particle detector  to discriminate the two options.
  \begin{figure}[t!]
\includegraphics[width=.49\textwidth,height=4cm]{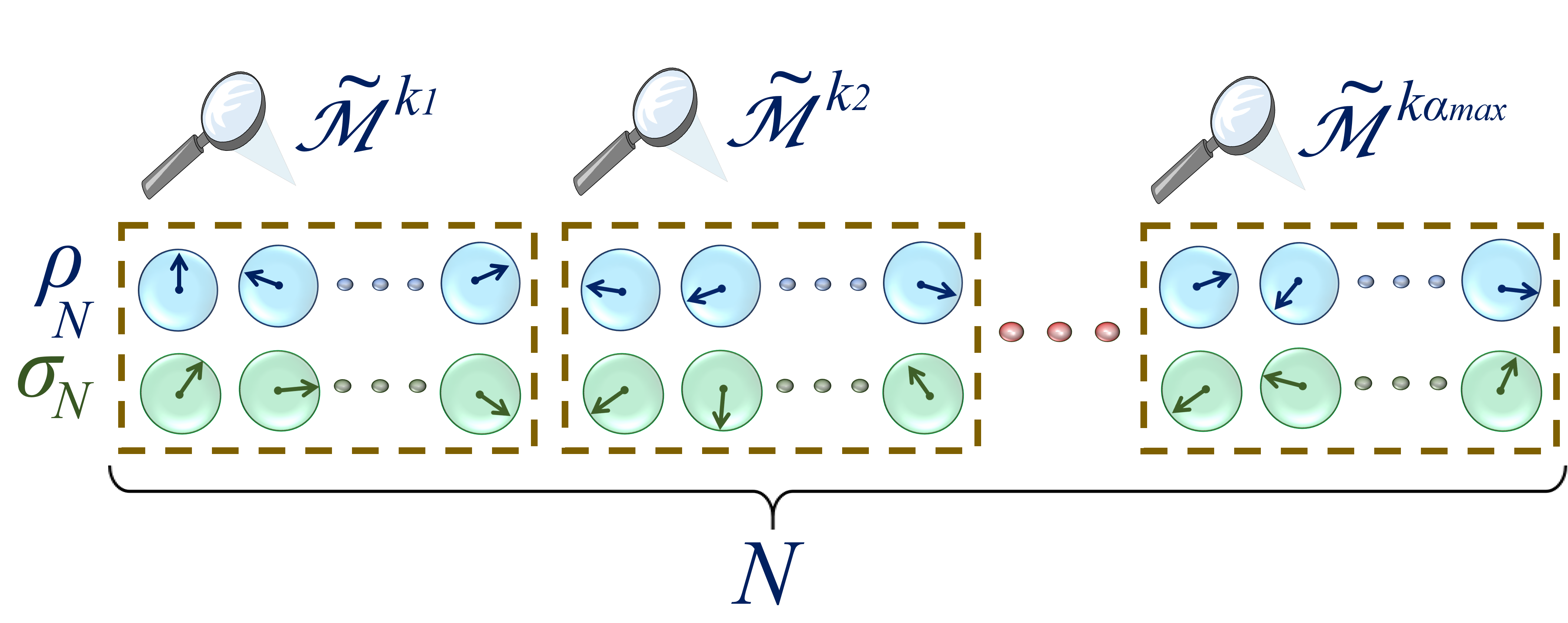}
\vspace{-20pt}
\caption{Consider  two $N$ particle states $\rho_N,\sigma_N$. A set of observers compute the distance between the marginal states of   subsystems with size  $k_\alpha,\,\sum_{\alpha}k_\alpha=N$, given by $d(\rho_{k_\alpha},\sigma_{k_\alpha})=\sum_i d_{cl}\left(\text{Tr}\left\{\tilde{{\cal M}}^{k_\alpha}_i \rho_{k_\alpha}\right\}, \text{Tr}\left\{\tilde{{\cal M}}^{k_\alpha}_i\sigma_{k_\alpha}\right\}\right)$. We quantify the difficulty to discriminate the two states by a weighted sum of each observer contribution.}
\label{fig0}
\end{figure}

\noindent Consider therefore a more general scenario, in which there is a set of cooperating observers that want to discriminate between $\rho_N$ and $\sigma_N$. Each of them performs the optimal  measurements $\tilde{{\cal M}}^{k_\alpha}$ to discriminate the states $\rho_{k_\alpha},\sigma_{k_\alpha}$ of subsystems composed of  $k_\alpha\leq N$ particles (\cref{fig0}), then computing $d(\rho_{k_\alpha},\sigma_{k_\alpha})$. The setup defines a measurement partition
  \begin{align*}
&P_{k_\alpha}:=\left\{\tilde{{\cal M}}^{k_\alpha},\,\sum_\alpha\,k_\alpha=N\right\}.
\end{align*}
 For example, given  $N=3$, there are the following options: three observers perform single-site detections, determining the partition  $\{\tilde{{\cal M}}^{1}, \tilde{{\cal M}}^{1}, \tilde{{\cal M}}^{1}\}$; an observer makes a bipartite measurement, and another one performs a single-particle measurement, inducing three possible partitions $\{\tilde{{\cal M}}^2,\tilde{{\cal M}}^1\}$ \cite{notepermutations}; a single observer implements a three-site measurement $\tilde{{\cal M}}^3$.  The measurements on different subsystems are independent and compatible, $\left[\tilde{{\cal M}}^{k_{\alpha_i}},\tilde{{\cal M}}^{k_{\alpha_j}}\right]=0,\,\forall\,\tilde{{\cal M}}^{k_{\alpha_i}},\tilde{{\cal M}}^{k_{\alpha_j}}\in P_{k_\alpha}$. Then, we might pick the sum of all the contributions, $\sum_{\alpha}d(\rho_{k_\alpha},\sigma_{k_\alpha})$, to quantify the  information that is extractable from  $P_{k_\alpha}$. %Note that it is crucial to consider sum of distances, in order to factor the particle number, because $d$ is generally not additive under tensor product.  
 Consequently,  the maximal value of the arithmetic sum over all the system partitions could be a new measure of state distinguishability. 
% over all possible partitions might be therefore a sound way to quantify state distinguishability.
 Unfortunately, this quantity would not take into account that each measurement is performed on a different number  of particles $k_\alpha$. It is experimentally harder to implement  $\tilde{{\cal M}}^k$ than any  $\tilde{{\cal M}}^{l<k}$. An extreme case is the discrimination of the GHZ  state from the classically correlated state $|a|^2\ket{0}\bra{0}^{\otimes N}+|b|^2\ket{1}\bra{1}^{\otimes N}$:  they are found to be identical by all measurement setups but a full scale $N$-particle detection. By increasing $N$, it becomes harder to distinguish the two preparations. Yet, the maximal distance sum  is  %$(a^4+b^4)^{1/2},\,\forall\,N$ 
$d(\rho_N,\sigma_N)$, which does not depend on $N$.  
%The contractivity of the distance function suggests that the more particle one monitors, the more informative is the measurement. 
 %We observe that the contractivity under partial trace of the distance function implies that for each cluster  of arbitrary size $k_\alpha$, one has $k_\alpha D(\rho_{k_\alpha},\sigma_{k_{\alpha}})\geq  \sum_{k_\beta\subset k_\alpha: \sum_\beta k_\beta= k_\alpha} k_{\beta}D(\rho_{k_\beta},\sigma_{k_\beta})$. \\
 A better choice is, for each partition $P_{k_\alpha}$, to  sum all the observer  contributions, while weighting their relative importance  by {\it the inverse of} the size of the measured subsystem:
 \begin{align}\label{eq3}
  \delta_{d,P_{k_\alpha}}(\rho_N,\sigma_N):=\sum_{\alpha}\frac{1}{k_\alpha}d\left(\rho_{k_\alpha},\sigma_{k_\alpha}\right).
 \end{align}
 This more refined quantity filters out system degeneracy, which manifests when two or more particles are in the same state. Comparing the two states $\rho_N=\ket{0}\bra{0}^{\otimes N}, \sigma_N=\ket{1}\bra{1}^{\otimes k}\ket{0}\bra{0}^{\otimes N-k}$, one has $\delta_{d,P_{k_\alpha}}(\rho_N,\sigma_N)\leq k\,M_d$. Note that, conversely, the weighted sum $\sum_{\alpha}k_\alpha d(\rho_{k_\alpha},\sigma_{k_\alpha})$  overvalues the difference between states. For example, by choosing the $N$ particle detection $\tilde{{\cal M}}^N$, one would have  $N\,d(\rho_N,\sigma_N)=N\,d(\rho_k,\sigma_k)=N\,M_d, \forall\,k$.\\ %In general, it can be increased by adding a redundant register of particles in the $\ket{0}$ state to  the system under study. 
We are now ready to quantify the ability to discriminate two arbitrary $N$-partite quantum states by a single index:\\

\noindent {\it We define the $d$ weighted distance between two states $\rho_N,\sigma_N$ as}
 \begin{align}\label{eq5}
 D_d(\rho_N,\sigma_N):=\max\limits_{P_{k_\alpha}} \delta_{d,P_{k_\alpha}}(\rho_N,\sigma_N).
 \end{align}
 
\noindent We further justify the definition. Since it is a (weighted) sum of distances with positive weights, the weighted distance  inherits the first, and fourth properties of the distance function in \cref{eq1}, which we listed in \cref{prop}. The second property, the faithfulness, is satisfied because it is the maximal one among all the weighted sums in \cref{eq3}. The third property, contractivity, holds for local CPTP maps performed on a  single subsystem. See the full proof in \cite{epaps}.
\begin{table*}
  \bgroup
\def\arraystretch{1.5}
\begin{tabular}{|c|c|c|}
\hline
  $\bm{\rho_N,\sigma_N}$& $\bm{B(\rho_N,\sigma_N) }$ & $\bm{D_B(\rho_N,\sigma_N)}$  \\
  \hline
 $\ket{0}^{\otimes N},\,\,\ket{1}^{\otimes k}\ket{0}^{\otimes N-k}$& $\frac\pi2,\,\forall\,k$ &     $k\,\frac\pi2$ \\
   \hline
  $\ket{0}^{\otimes N},\,\,\ket{ghz_k}\otimes\ket{0}^{\otimes N-k }$ & $\cos^{-1} |a|$ &  $k\,\cos^{-1}|a|$    \\
  \hline
  $\ket{0}^{\otimes N},\,\,\ket{ghz_l}^{\otimes k}\ket{0}^{\otimes N-k\,l}$&$ \cos^{-1}|a|^k,\,\forall\,l$ &     $k\,l \cos^{-1}|a|$ \\
\hline
 $\ket{0}\bra{0}^{\otimes N}, class_k\otimes \ket{0}\bra{0}^{\otimes N-k}$& $\cos^{-1}|a|$ &  $k\,\cos^{-1}|a|$    \\
 \hline 
 $\ket{0}\bra{0}^{\otimes N}, class_l^{\otimes k}\otimes \ket{0}\bra{0}^{\otimes N-k\;l}$& $\cos^{-1} |a|^k,\,\forall\,l$ &  $k\,l\,\cos^{-1}|a|$    \\
   \hline
    $\ket{0}^{\otimes N}, \ket{dicke_{N,k}}$& $\frac\pi2,\,\forall\,k$  &  $N \cos^{-1}\left(1-\frac kN\right)$    \\
 \hline
     $\ket{0}\bra{0}^{\otimes N},\,I_k/2^k\otimes \ket{0}\bra{0}^{\otimes N-k}$& $\cos^{-1} \frac1{\sqrt{2^k}}$ &     $k \cos^{-1}\frac{1}{\sqrt 2}$ \\
   \hline
  $\ket{ghz_N}\bra{ghz_N},\, I_N/2^N,\,|a|,|b|\neq\frac{1}{\sqrt2}$& $\cos^{-1}\left(\frac{|a|+|b|}{\sqrt{2^N}}\right)$ & $N\,\cos^{-1}\left(\frac{|a|+|b|}{\sqrt2}\right)$       \\
  \hline
  $class_N,\,I_N/2^N,\,|a|,|b|\neq\frac{1}{\sqrt2}$&$\cos^{-1}\left(\frac{|a|+|b|}{\sqrt{2^N}}\right)$   &      $N\cos^{-1}\left(\frac{|a|+|b|}{\sqrt2}\right)$    \\
   \hline
  $\ket{ghz_N}\bra{ghz_N},\,I_N/2^N,\,
  N\,\text{even},\, |a|=|b|=\frac{1}{\sqrt2}$ & $\cos^{-1}\frac{1}{\sqrt{2^{N-1}}}$ & $\frac{N\pi}{16}$       \\
 \hline
  $class_N,\,I_N/2^N,\,
    N\,\text{even},\, |a|=|b|=\frac{1}{\sqrt2}$&$\cos^{-1}\frac{1}{\sqrt{2^{N-1}}}$ &     $\frac{N\pi}{16}$  \\
     \hline 
 $class_N,\, \ket{ghz_N}\bra{ghz_N}$&$\cos^{-1}\sqrt{a^4+b^4}$&$\frac{\cos^{-1}\sqrt{a^4+b^4}}{N}$\\
 \hline
\end{tabular}
 \egroup
\caption{We calculate the standard Bures length and  the  weighted Bures length, as defined in \cref{bweight},   for $N$ qubit states (full details in \cite{epaps}). Here $\ket{ghz_k}=\left(a\ket{0}^{\otimes k}+b\ket{1}^{\otimes k}\right)$, $class_k=\left(|a|^2\ket{0}\bra{0}^{\otimes k}+|b|^2\ket{1}\bra{1}^{\otimes k}\right)$, and $\ket{dicke_{N,k}}=\frac{1}{\sqrt{\binom{N}{k}}}\sum_i{\cal P}_i\ket{0}^{\otimes N-k}\ket{1}^{\otimes k}$ is the $N$ qubit Dicke state with $k$ excitations \cite{dicke}, in which ${\cal P}_i$ are the possible permutations. The weighted Bures length is a better descriptor of the difference between multipartite quantum states. If two states become more different by increasing $N$, i.e. there are more measurement setups that discriminate between them, the quantity increases. If    discriminating  two states becomes harder,  the  weighted Bures length decreases.  
%The quantity outperforms the standard distance even in discriminating between a pair of classical states. %Note that  $class_k$ and   $\ket{ghz_k}$ are equally far from an input state $\ket{0}^{\otimes N}$. The result is expected, as we are measuring state distinguishability without, for the moment
}\label{tab1}
\end{table*}
 \noindent The weighted distance is invariant only under single  particle unitary maps, while the standard distance $d$ is invariant under all unitaries. This property is crucial for comparing  many-body configurations, capturing the fact that the states $\ket{00},a\ket{00}+b\ket{11}$ are more different than $\ket{00}, a\ket{00}+b\ket{10}$. The weighted distance is bounded via the chain of inequalities%The weighted distance is upper bounded by the total information that is stored in the system: 
 \begin{align}\label{bound}
 \frac{1}{N}\,d(\rho_N,\sigma_N)\leq D_d(\rho_N,\sigma_N)\leq N\,d(\rho_N,\sigma_N)\leq N\,M_d,
 \end{align}
 %in which we assumed that the distance function is normalized to have maximal value equal to one.
being  maximal for ``maximally different'' preparations,  such that   both the global states and all their marginal states are orthogonal. Note that the importance of the largest measurement setup does not increase under trivial extensions of the system.  For example, consider the $N$-partite states $\ket{0}^{\otimes N}, \ket{x_1x_2\ldots x_N}$. By adding a $Q$-particle register in $\ket{0}^{\otimes Q}$, the new states are $\ket{0}^{\otimes N+Q}, \ket{x_1x_2\ldots x_N}\ket{0}^{\otimes Q}$. One has $(N+Q)\,d(\rho_{N+Q},\sigma_{N+Q})\geq N\,d(\rho_N,\sigma_N)$, while  $D_{d}(\rho_{N+Q},\sigma_{N+Q})=D_{d}(\rho_N,\sigma_N)$, since an $N$-particle detection ${\tilde {\cal M}}^N$ is still maximally informative.\\  %One may object that the dimension of particle  should be taken into account too. %Discriminating between preparations of an $n$-dimensional single system $\ket{0}, a\ket{0}+b\ket{n-1}$ should be deemed to be more difficult than distinguishing between  $\ket{0}, a\ket{0}+b\ket{1}$. 
 %Note that  the information about an $n$-dimensional  system can be encoded in $N$ qubits, given $2^N>n$. Hence, we focus on the multiqubit case with no loss of generality.\\  %: ${\cal D}_D(\rho_N, \sigm}_N)=N\,D(\tilde\rho_N, \tilde{\sigma}_N)$.  
 
\noindent We test the usefulness of the notion of weighted distance. Adopting as standard distance the Bures length $B(\rho_N,\sigma_N):=\cos^{-1} F(\rho_N,\sigma_N)$ \cite{uhl,bures,notebures}, motivated by the considerations  detailed via \cref{eq1,prop,eq3,eq5}, we define the weighted Bures length: 
\begin{align}\label{bweight}
 D_{B}(\rho_N,\sigma_N):=\max\limits_{P_{k_\alpha}} \delta_{B,P_{k_\alpha}}(\rho_N,\sigma_N).
\end{align}
 We compare the two quantities via explicit calculations in some interesting case studies, see \Cref{tab1}. 
  The results confirm that the weighted Bures length is  more informative than the standard Bures length. For pure states, the latter is equal to the Fubini-Study distance \cite{brau}. Consequently, \cref{bweight} defines a weighted Fubini-Study distance for pure states. %We can therefore build a weighted statistical distance for factorized wavefunctions $\otimes_{i=1}^N\ket{i}$ as Non-negativity and faithfulness are met by construction.
 In general, the full knowledge of the quantum states under study is required for exact calculations of both standard and weighted distances, but statistical methods for estimating standard distances from incomplete data are readily applicable, by construction, to weighted distance estimation  \cite{flammia,overlap,fido}. \\ % \\The maximization in \cref{bweight} runs over all the possible partitions, making the evaluation of the weighted distance potentially hard for states without evident symmetries, e.g. invariance under particle permutations. However, computationally friendly lower bounds can be built by maximizing \cref{eq3} over a certain subset of partitions. \\

\begin{figure}[t!]
\includegraphics[width=.4\textwidth,height=5cm]{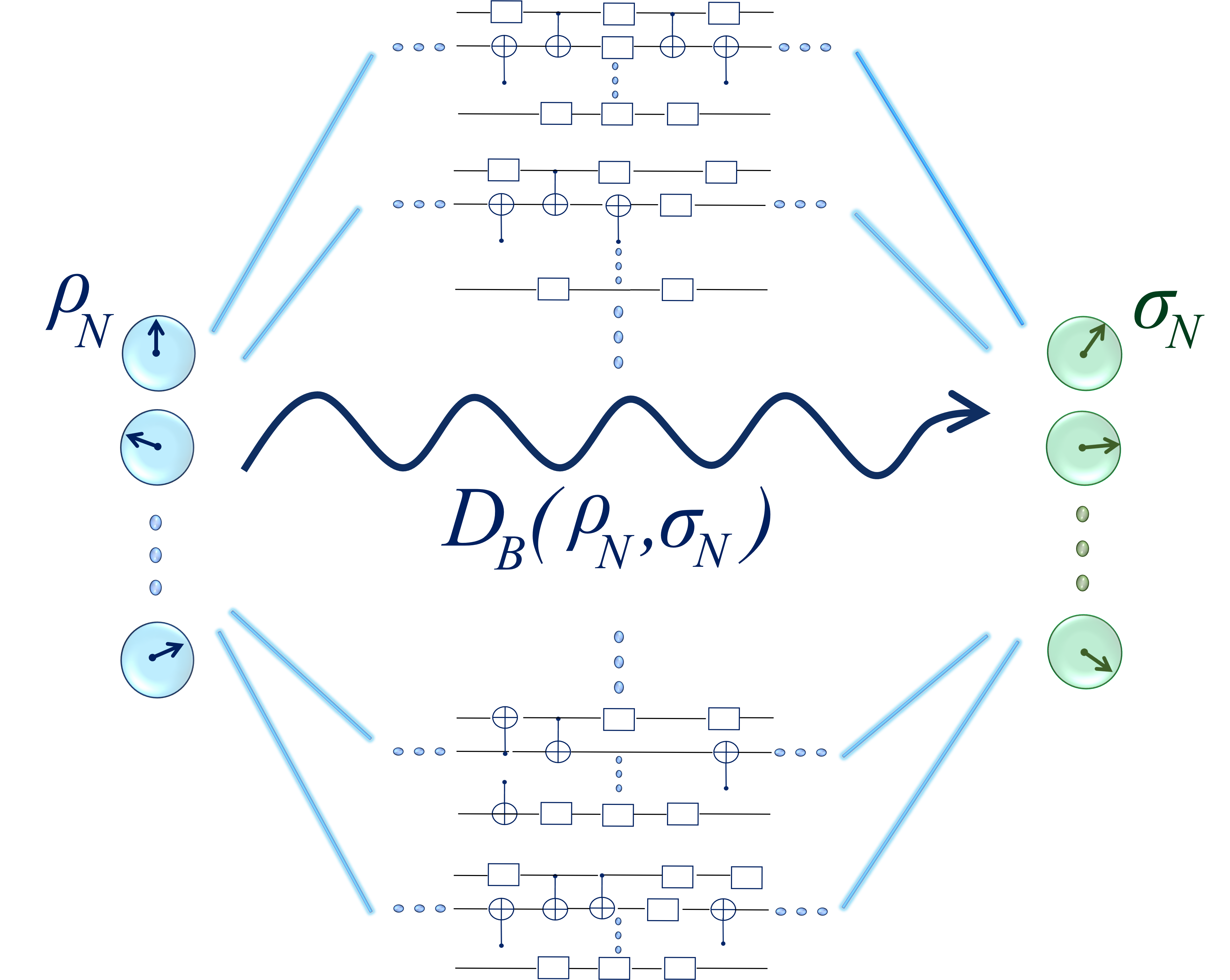}
\caption{We prove that the weighted Bures length $D_B(\rho_N,\sigma_N)$ is a lower bound to the experimental cost of the state transformation $\rho_N\rightarrow \sigma_N$.  The bound is also valid for non-unitary quantum processes.}
\label{fig}
\end{figure}

\noindent {\it The weighted Bures length lower bounds the experimental cost of quantum processes --} The weighted distances have a clear metrological meaning, being more sophisticated proxies  than standard distances for state discrimination \cite{metrology}. An important related question is what is the cost of creating very different configurations in terms of physical resources, such as energy and time. Specifically, generating highly correlated states from  $\ket{0}^{\otimes N}$,  transforming an initial state in a very different output,  is a requisite of all quantum algorithms. Establishing the physical limits to  quantum programming, i.e. how small state preparation circuits can be, is therefore of great interest, as environmental noise quickly corrupts them \cite{open}.  The results in \Cref{tab1} highlight that, when calculated between an initial state $\ket{0}^{\otimes N}$ and highly correlated outputs, the weighted Bures length is monotonically increasing  with the size of the system.  We show that, indeed, the  weighted Bures length between the initial and final states of a quantum process is the minimum experimental cost of the state transformation. We employ a geometric argument to rigorously prove the claim (\cref{fig}).
\\

\noindent A quantum dynamics from an $N$-qubit input state $\rho_{N}$ to a final state $\sigma_N$ is a path in the stratified Riemannian manifold of density matrices \cite{amari,geo1}. The state of the system at time $t$ has spectral decomposition $\rho_{N,t}=\sum_{r=1}^{2^N}\lambda_r(t)\ket{r(t)}\bra{r(t)},\,t\in[0,T]$, with $\rho_{N,0}\equiv \rho_{N},\,\rho_{N,T}\equiv\sigma_N$. Its rate of change is the time derivative $\dot\rho_{N,t}$. One builds a distance measure between two  quantum states $\rho_{N},\sigma_N$ by calculating the minimum of the length functional $\int_{0}^{T} ||\dot\rho_{N,t}||\,dt$ for some given norm. In particular, the input/output Bures length is the distance induced by the Fisher norm \cite{petz}:
\begin{align}\label{fisher}
B(\rho_{N},\sigma_N)&=\min\limits_{\rho_{N,t}}\int_{0}^{T} ||\dot\rho_{N,t}||_{\cal F}\,dt,\nonumber\\
 ||\dot\rho_{N,t}||^2_{{\cal F}} &\coloneqq \sum_{r}\frac{\dot\lambda_r^2(t)}{4\,\lambda_r(t)}+\sum_{r<s }\frac{|\bra{r(t)}\dot\rho_{N,t}\ket{s(t)}|^2}{\lambda_r(t)+\lambda_s(t)}.
\end{align}
 The first term  in  \cref{fisher} is the classical Fisher norm. The second one is  a purely quantum contribution (related to the state eigenbasis evolution), being the only term surviving for unitary maps (the two terms coexist for generic CPTP operations). We  evaluate the cost of eigenbasis changes, adopting the viewpoint that classical computations are free. The transformation can be split into two steps: the eigenvalue change and the eigenbasis change: $\rho_N\rightarrow \tau_N\rightarrow \sigma_N$, in which $\tau_N=\sum_{r=1}^{2N}\lambda_r(T)\ket{r(0)}\bra{r(0)}$ \cite{noteiso}.  The first step can be always completed via a classical process \cite{me}, while the second one can be implemented by a unitary path $\tau_{N,t},\,\tau_{N,0}\equiv\tau_N, \tau_{N,T}\equiv\sigma_N$. For unitary processes, the first step is redundant, $\rho_N=\tau_N$. Hence, we quantify the ``quantum cost'' for implementing an arbitrary (even non-unitary) transformation $\rho_{N}\rightarrow \sigma_N$ as 
\begin{align}
B^q(\rho_N,\sigma_N):=\min\limits_{\text{unitary paths}\,\tau_{N,t}}\int_{0}^{T} ||\dot\tau_{N,t}||_{\cal F}\,dt. 
 \end{align}
   Suppose we carry out the second step via a sequence of quantum gates  $U= \Pi_l U_l, U_l=e^{-i\,H_l\,T_l}$ (we run $U_1$, then $U_2$, and so on).  The spectral decomposition of each time-independent Hamiltonian is $H_l=\sum_{x_l=1}^{2^{k_l}} h_{x_l}\ket{h_{x_l}}\bra{h_{x_l}}, \,h_{x_{l>m}}\geq h_{x_m}, \forall\,l,m,$ and $T_l$ is the runtime of each gate. Note that any Hamiltonian $H_l$ affects $k_l\leq N$ particles.  Call $\tau^l_{N,t_l}$ the intermediate state at time $t_l\in [0,T_l]$  while implementing $U_l$, with $ \tau^l_{N, 0}\equiv \tau^l_N, \tau^{1}_{N,0}\equiv\tau_N$. Since time-independent Hamiltonian dynamics are constant speed processes, one has
\begin{align}
B^q(\rho_N,\sigma_N)&\leq \int_{0}^{\sum_lT_l} ||\dot{\tau}_{N,t}||_{{\cal F}}\,dt \\\nonumber
&=\sum_l\int_{0}^{T_l} \left|\left|\dot{\tau}_{N,t_l}^l\right|\right|_{{\cal F}}\,dt_l=\sum_l||\dot{\tau}^l_N||_{{\cal F}}\, T_l.
%&\geq \min\limits_{\rho_t} \int_0^T\sum_{r<s }\frac{|\bra{r(t)}\dot\rho_t\ket{s(t)}|^2}{\lambda_r(t)+\lambda_s(t)}\,d\,t,
\end{align}
   The inequality can be saturated when  $\sigma_N$ (and therefore $\tau_N$) is a pure state. %When the classical step is unnecessary, $\rho=\tilde{\rho}$. % , one has $d^q_B(\rho_{in},\rho_f)$. 
The squared speed of the process lower bounds  the variance of the generating Hamiltonian, which is also constant in time \cite{toth}:
\begin{align}
V_{\tau^l_N}(H_l):=\text{Tr}\left\{H_l^2\,\tau^l_N\right\}-\text{Tr}\left\{H_l\,\tau^l_N\right\}^2\geq ||\dot{\tau}^l_N||_{{\cal F}}^2,\, \forall\,l.
%\text{Tr}\left\{H^2\,\tilde\rho_{t,U}\right\}-\text{Tr}\left\{H\,\tilde\rho_{t,U}\right\}^2\geq ||\dot{\tilde{\rho}}_{U}||_{{\cal F}}^2. %\left(d_B^q(\rho_{in},\rho_f)/T\right)^2,\,\forall t.
\end{align}
  By employing the  (halved) semi-norm $E_l:=\left(h_{x_l=2^{k_{l}}}-h_{x_l=1}\right)/2$ \cite{semi}, we quantify the cost of the state transformation  in terms of physical resources by
\begin{equation}
{\cal R}_{U_l}:=k_l\,E_l\,T_l \Rightarrow {\cal R}_U:=\sum_l {\cal R}_{U_l}.
\end{equation}
The first term $k_l$ represents the size of each quantum gate $U_l$.  The second term  quantifies the energy requirement for each gate. Note that  $E_l^2\geq V_{\rho_l}(H_l),\,\forall\,l$.  The third contribution is the allowed time interval for each gate. Factoring in the gate size is essential. A single qubit Hamiltonian  of spectrum $(x,-x)$ is easier to implement, in some given time $T_l$, than a $k>1$-partite interaction generated by $(\underbrace{x,0,\ldots,-x}_{2^k})$, even though the eigenvalue gap $E_l$ is equal.  %In general, the experimental cost of a quantum gate significantly depends on the number of affected particles. 
 By remembering \cref{bound}, and exploiting the triangle inequality of the weighted distances, it follows that\\
 
 \noindent {\it The experimental cost ${\cal R}_{U}$ of a state transformation $\rho_N\rightarrow \sigma_N$ is lower bounded by the   weighted Bures length between initial and final states:}
 \begin{align}\label{buresbound}
& {\cal R}_{U_l}\geq k_l \, B^q(\tau^{l}_N,\tau^{l+1}_N) \geq D_{B}(\tau^l_N,\tau^{l+1}_N),\,\forall\,l\Rightarrow\nonumber\\
&\text{For unitary processes:}\,\, {\cal R}_U \geq D_B(\rho_N,\sigma_N)\\
&\text{For general quantum processes:}\,\, {\cal R}_U \geq D_B(\tau_N,\sigma_N).\nonumber
 \end{align}
 The bounds are formally similar  to energy-time uncertainty relations and quantum speed limits \cite{speed,unc,luo,gibi,speed2,speed3}, yet they can be more informative, as they provide a more nuanced  resource count for quantum processes. For example, they determine the minimum time to complete state transformations at fixed energy and gate size. %The left-hand side of \cref{buresbound} is a product of the available energy and time for the process, but the size of the interacting Hamiltonians $H_l$ is also factored.
  Note that the right-hand side is zero if and only if $[\rho_N,\sigma_N]=0$. That is, if and only if there exists a classical dynamics that transforms the input into the output state \cite{me}. The left-hand inequality in \cref{buresbound} is saturated when the intermediate states $\tau^{l}_N$ are the most sensitive ones to the unitary perturbations $U_l$, i.e. they are coherent superpositions $\left(\ket{h_{2^{k_l}}}+e^{i\,\phi}\ket{h_{x_l=1}}\right)/\sqrt 2,\,\phi\in[0,2\,\pi]$. %Note that there is actually a more general bound for time-independent processes \begin{\align} {\cal R}_{U} \end{\align}
%The related distance function is the Bures distance $D_B(\rho_i,\rho_f)=\sqrt{2-2F(\rho_i,\rho_f)},$  in which $F(\rho_i,\rho_f)=\text{Tr}\left\{\sqrt{\sqrt{\rho_f}\rho_i\sqrt{\rho_f}}\right\}$ is the fidelity, which is equal to the state overlap between pure states.  
%The operational meaning of the fidelity makes clear why it cannot capture the difference between states of many-body quantum systems. The  fidelity evaluates how difficult it is to distinguish  two states via measurements. Yet, it does not take in account the size of the optimal measurement apparatus. The problem is shared by other popular metrics, such as the trace distance and the Hilbert-Schmidt distance, and by all their classical counterparts.\\
The result in \cref{buresbound} advances our understanding of many-body quantum processes in three ways.  First, it provides a lower limit to the difficulty to run  quantum computations in terms of an exact, analytical bound, rather than an order of magnitude \cite{aar,susk,nielsengeo}. Second, it applies to mixed states and non-unitary processes, beyond the idealized scenario of  perfectly controllable quantum dynamics. Third, the right-hand side of the bound, the  weighted Bures distance, is not just a numerical value, but it has a physical meaning. Specifically, the bound highlights that our ability to manipulate quantum states, e.g. generating entangled configurations from the input state $\ket{0}^{\otimes N}$, is never greater than the instrumental experimental cost.\\
%the relative  state complexity is a metric for evaluating the size of  the shortest algorithm (up to lower order corrections) to prepare a quantum state from a given input \cite{aar}. While the state complexity can in fact be linked to the geometry of quantum processes \cite{susk,nielsengeo}, the concept does not straightforwardly apply to mixed states and non-unitary evolutions.   
%The same limitation affects  the quantum volume \cite{vol}, another popular performance index for quantum devices.\\
 
\noindent{\it Conclusion. -- } We have introduced the weighted distances (\cref{eq5}), a new class of information measures. They capture the difficulty  in distinguishing many-body quantum states.  % The weighted distances are justified by proving that they satisfy a set of desirable properties. 
Moreover, we uncovered a fundamental bound to quantum information processing (\cref{buresbound}). The size of state preparation algorithms is never smaller than the  weighted Bures length between the input and the output states, i.e. our ability to discriminate between the two states. %Quantifying the difference between two quantum configurations is a key goal of quantum theory, as well as a crucial step in the diagnosis of quantum experiments. 
 %The quest is motivated by experimental reasons too.  Near-term quantum devices are expected to be very noisy. %Both predicting their computational power and evaluating their performance  requires building error-aware quantum algorithms which replace idealized pure states and unitary evolutions with mixed states and general quantum operations.
   We  anticipate that  the weighted distances will help evaluate the theoretical and experimental performance of quantum technologies \cite{vol}, and explore critical properties of open quantum systems \cite{delgado}.  \\
  % Follow-on studies may verify that weighted state and gate distances are   inherently more informative than the currently popular metrics about the efficiency of state preparation and manipulation.  \\

 \noindent{\it Acknowledgments. --}
We thank Pawel Kurzynski for pointing out that the contractivity of the weighted distances holds only for single site operations, and Andrey Bagrov, Tom Westerhout, and an anonymous Referee for other useful comments.  The research presented in this  article was supported by a Rita Levi Montalcini Fellowship of the Italian Ministry of Research and Education (MIUR), grant number 54$\_$AI20GD01, and by the  Templeton World Charity Foundation Power of
Information Fellowship.

\clearpage
\onecolumngrid
\renewcommand{\bibnumfmt}[1]{[A#1]}
% citenumfont command adds A to all numbers
\renewcommand{\citenumfont}[1]{{A#1}}

\setcounter{page}{1}
\setcounter{equation}{0}
 
\appendix*
\section{Supplementary Material}
  \subsection{Proof that the weighted distances satisfy the properties of standard distances}
  We prove that the weighted distances defined in eq.~4 of the main text meet the properties that are listed in eq.~2.
  \begin{itemize}
 \item $D_d(\rho_N,\sigma_N) \geq 0$ (non-negativity):\\
 Since the weighted distance is a weighted sum of  standard distances, i.e. non-negative quantities, and the weights $\frac{1}{k_\alpha}$ are positive, the property is satisfied.
 \item $D_d(\rho_N,\sigma_N)=0\iff \rho_N=\sigma_N$ ( faithfulness):\\
 Consider all the possible weighted sums defined in eq.~3 of the main text. Since the weights are positive, if and only if there is at least a non-zero term $d(\rho_{k_{\alpha}},\sigma_{k_\alpha})$, there is a non-zero weighted sum $\delta_{d,P_{k_\alpha}}(\rho_N,\sigma_N)$. Since the weighted distance is the maximal weighted sum, the claim is proven.
 \item $D_d(\rho_N,\sigma_N)\geq D_d(\Lambda (\rho_N),\Lambda (\sigma_N)),\,\forall\,\Lambda,$ (contractivity for CPTP maps on a single subsystem):\\
 Since the weighted distance is a weighted sum of  standard distances, i.e. contractive quantities, and the weights $\frac{1}{k_\alpha}$ are positive, the property is satisfied.
 \item $D_d(\rho_N,\sigma_N)\leq  D_d(\rho_N,\tau_N)+D_d(\tau_N, \sigma_N)$ (triangle inequality):\\
 For any term of a weighted sum, one has $\frac{1}{k_\alpha}d\left(\rho_{k_\alpha},\sigma_{k_\alpha}\right)\leq \frac{1}{k_\alpha} \left(d\left(\rho_{k_\alpha},\tau_{k_\alpha}\right)+d\left(\tau_{k_\alpha},\sigma_{k_\alpha}\right)\right)$. Consider now the partition $\hat P_{k_\alpha}$  that defines the maximal weighted sum, i.e. the weighted distance, between $\rho_N$ and $\sigma_N$. One has 
 \begin{align*}
 D_d(\rho_N,\sigma_N)= \delta_{d,\hat P_{k_\alpha}}(\rho_N,\sigma_N) =\sum_{\alpha}\frac{1}{\hat k_\alpha}d\left(\rho_{\hat k_\alpha},\sigma_{\hat k_\alpha}\right)\leq \sum_{\alpha}\frac{1}{\hat k_\alpha}\left(d\left(\rho_{\hat k_\alpha},\tau_{\hat k_\alpha}\right)+d\left(\tau_{\hat k_\alpha},\sigma_{\hat k_\alpha}\right)\right)\leq D_d(\rho_N,\tau_N)+D_d(\tau_N,\sigma_N).
 \end{align*}
 
 \end{itemize}
 \subsection{Calculation of weighted Bures length in some interesting cases}
    
 \noindent Here we provide details about the results  reported in Table 1 of the main text. We calculate the Bures length and the weighted Bures length for several pairs of $N$ qubit states $\rho_N,\sigma_N$. The two quantities respectively read:
 \begin{align*}
 B(\rho_N,\sigma_N)&:=\cos^{-1}F(\rho_N,\sigma_N),\\
 D_{B}(\rho_N,\sigma_N)&:=\delta_{B,N}(\rho_N,\sigma_N)= \max\limits_{P_{k_\alpha}}\sum_\alpha\frac{1}{k_\alpha}B(\rho_{k_\alpha},\sigma_{k_\alpha}). 
 \end{align*}
 The standard Bures length is straightforwardly computed from the global density matrices under study. The weighted Bures length was obtained for each case in the Table as follows.
 \begin{itemize}
 \item Single particle measurements, defining the partition $\{\underbrace{\tilde{\cal M}^1, \tilde{\cal M}^1,\,\ldots, \tilde{\cal M}^1}_{N}\},$ are sufficient to discriminate  between $\ket{0}^{\otimes N}$ and any other state. Hence, the weighted Bures length is the sum of the Bures length of single particle density matrices:
 \begin{itemize}[leftmargin=.5em]
 \item[$\star$] $\rho_N=\ket{0}\bra{0}^{\otimes N},\,\sigma_N=\ket{1}\bra{1}^{\otimes k}\ket{0}\bra{0}^{\otimes N-k}\Rightarrow D_{B}(\rho_N,\sigma_N)=k\,\cos^{-1}|\langle0|1\rangle|=k\,\frac\pi2$
 \item[$\star$] $\rho_N=\ket{0}\bra{0}^{\otimes N},\,\sigma_N=\ket{ghz_k}\bra{ghz_k}\otimes\ket{0}\bra{0}^{\otimes N-k }\Rightarrow  D_{B}(\rho_N,\sigma_N)=k\,\cos^{-1}\text{Tr}\left|\ket 0\bra{0}\left(|a|\ket{0}\bra{0}+|b|\ket{1}\bra{1}\right)\right|=k\,\cos^{-1}|a|$
 
 \item[$\star$] $\rho_N=\ket{0}\bra{0}^{\otimes N},\,\sigma_N=\ket{ghz_l}\bra{ghz_l}^{\otimes k}\ket{0}\bra{0}^{\otimes N-k\,l}\Rightarrow D_{B}(\rho_N,\sigma_N)=$\\
 $\underbrace{k\,\cos^{-1}\text{Tr}\left|\ket 0\bra{0}\left(|a|\ket{0}\bra{0}+|b|\ket{1}\bra{1}\right)\right|+\ldots+k\,\cos^{-1}\text{Tr}\left|\ket 0\bra{0}\left(|a|\ket{0}\bra{0}+|b|\ket{1}\bra{1}\right)\right|}_{l}=k\,l \cos^{-1}|a|$

 \item[$\star$] $\rho_N=\ket{0}\bra{0}^{\otimes N}, \sigma_N=class_k\otimes \ket{0}\bra{0}^{\otimes N-k}  \Rightarrow  D_{B}(\rho_N,\sigma_N)=k\,\cos^{-1}\text{Tr}\left|\ket 0\bra{0}\left(|a|\ket{0}\bra{0}+|b|\ket{1}\bra{1}\right)\right| =k\,\cos^{-1}|a|$

 \item[$\star$]$\rho_N=\ket{0}\bra{0}^{\otimes N}, \,\sigma_N=class_l^{\otimes k}\otimes \ket{0}\bra{0}^{\otimes N-k\;l} \Rightarrow D_{B}(\rho_N,\sigma_N)=$\\
 $\underbrace{k\,\cos^{-1}\text{Tr}\left|\ket 0\bra{0}\left(|a|\ket{0}\bra{0}+|b|\ket{1}\bra{1}\right)\right|+\ldots+k\,\cos^{-1}\text{Tr}\left|\ket 0\bra{0}\left(|a|\ket{0}\bra{0}+|b|\ket{1}\bra{1}\right)\right|}_{l}=k\,l \cos^{-1}|a|$ 
 \item[$\star$]$\rho_N=\ket{0}\bra{0}^{\otimes N},\,\sigma_N= \ket{dicke_{N,k}}\bra{dicke_{N,k}}\Rightarrow  D_{B}(\rho_N,\sigma_N)=N\, \cos^{-1}\left|\ket{0}\bra{0}\left(\frac{\binom{N-1}{k}}{\binom{N}{k}}\ket{0}\bra{0}+\frac{\binom{N-1}{k-1}}{\binom{N}{k}}\ket{1}\bra{1}\right)\right|=N \cos^{-1}\left(1-\frac kN\right)$
 \item[$\star$] $\rho_N=\ket{0}\bra{0}^{\otimes N},\,\sigma_N=I_k/2^k\otimes \ket{0}\bra{0}^{\otimes N-k}\Rightarrow  D_{B}(\rho_N,\sigma_N)=k\,\cos^{-1}\text{Tr}\left|\ket 0\bra{0}\left(\frac{\ket{0}\bra{0}+\ket{1}\bra{1}}{\sqrt2}\right)\right|=k \cos^{-1}\frac{1}{\sqrt 2}$
 \end{itemize}
 \item Single-site detections are also sufficient to discriminate the correlated states $class_N, ghz_N$ from the identity, for $|a|,|b|\neq \frac1{\sqrt2}$:
 \begin{itemize}
  \item[$\star$] $\rho_N=\ket{ghz_N}\bra{ghz_N},\,\sigma_N=I_N/2^N,\,N\,\text{even}\Rightarrow  D_{B}(\rho_N,\sigma_N)=N\,\cos^{-1}\left(\frac{|a|+|b|}{\sqrt2}\right)$
 \item[$\star$]$\rho_N=class_N,\,\sigma_N=I_N/2^N,\,N\,\text{even}\Rightarrow  D_{B}(\rho_N,\sigma_N)=N\,\cos^{-1}\left(\frac{|a|+|b|}{\sqrt2}\right)$
 \end{itemize}
In case the state displays maximal classical or quantum correlations, i.e. $|a|=|b|=1/\sqrt2$, the best discriminating measurement setups are, for $N$ even, two-particle detections $\left\{\underbrace{\tilde{{\cal M}}^2, \tilde{{\cal M}}^2,\ldots, \tilde{{\cal M}}^2}_{N/2}\right\}$:
 \begin{itemize}
  \item[$\star$] $\rho_N=\Big|ghz_{N}\Big\rangle\Big\langle ghz_{N}\Big|,\,\sigma_N=I_N/2^N,\,N\,\text{even},\,|a|=|b|=1/\sqrt2\Rightarrow  D_{B}(\rho_N,\sigma_N)=\frac N4\,\cos^{-1}\left(\frac1{\sqrt2}\right)=\frac{N\,\pi}{16}$
 \item[$\star$]$\rho_N=class_{N},\,\sigma_N=I_N/2^N,\,N\,\text{even},\,|a|=|b|=1/\sqrt2\Rightarrow  D_{B}(\rho_N,\sigma_N)=\frac N4\,\cos^{-1}\left(\frac1{\sqrt2}\right)=\frac{N\,\pi}{16}$
 \end{itemize}
 \item The classical and quantum correlated states can be distinguished only by a full-scale measurement $\tilde{{\cal M}}^N$:
% \vspace{-5pt}
 \begin{itemize}
 \item[$\star$] $\rho_N=class_N,\,\sigma_N=\ket{ghz_N}\bra{ghz_N}\Rightarrow  D_{B}(\rho_N,\sigma_N)=\frac1N\,B(\rho_N,\sigma_N)=\frac{\cos^{-1}\sqrt{a^4+b^4}}{N}$.
 \end{itemize}
 \end{itemize}


\begin{thebibliography}{99}

\bibitem{wootters}W. K. Wootters, Statistical distance and Hilbert space, Phys. Rev. D 23, 357 (1981).

\bibitem{fubini}G. Fubini, Sulle metriche definite da una forme Hermitiana,  Atti del Reale Istituto Veneto di Scienze, Lettere ed Arti 63,  502 (1904).
\bibitem{study}E. Study, Kürzeste Wege im komplexen Gebiet. Mathematische Annalen, Springer Science and Business Media LLC. 60, 321 (1905).


  \bibitem{ghz}D. M. Greenberger, M. A. Horne, and A. Zeilinger, Going Beyond Bell's Theorem,  ``Bell's Theorem, Quantum Theory, and Conceptions of the Universe'', M. Kafatos (Ed.), Kluwer, Dordrecht, 69 (1989); arXiv:0712.0921.

  \bibitem{nielsen}M. A. Nielsen and I. L. Chuang, Quantum Computation and Quantum Information, Cambridge University Press (2000).


 \bibitem{fidelity1}C. A. Fuchs and C. M. Caves, Ensemble-Dependent Bounds for Accessible Information in Quantum Mechanics, Phys. Rev. Lett. 73, 3047(1994).
 \bibitem{fidelity2}R. Jozsa, Fidelity for Mixed Quantum States, J. Mod. Opt. 41, 2315 (1994). 
\bibitem{geo1}I. Bengtsson and K. Zyczkowski, Geometry of Quantum States, Cambridge University Press, Cambridge (2007).

\bibitem{preskill}J. Preskill, Quantum Computing in the NISQ era and beyond, Quantum 2, 79 (2018).
 \bibitem{fidelity3}Y.-C. Liang, Y.-H. Yeh, P. E. M. F. Mendonça, R. Y. Teh, M. D. Reid, and P. D. Drummond, Quantum fidelity measures for mixed states, Rep. Prog. Phys. 82, 076001 (2019).
\bibitem{fido1}E. Knill, D. Leibfried, R. Reichle, J. Britton, R.B.
Blakestad, J. D. Jost, C. Langer, R. Ozeri, S. Seidelin,
and D. J. Wineland, Randomized benchmarking of
quantum gates. Phys. Rev. A, 77, 012307 (2008).
\bibitem{fido2}T. Monz, P. Schindler, J. T. Barreiro, M. Chwalla, D. Nigg, W. A. Coish, M. Harlander, W. H\"{a}nsel, M. Hennrich,
and R. Blatt, 14-qubit entanglement: Creation and
coherence Phys. Rev. Lett., 106, 130506 (2011).
\bibitem{fido3}P. Sekatski, J.-D. Bancal, S. Wagner, and N. Sangouard, Certifying the building blocks of quantum computers from Bell's theorem, Phys. Rev. Lett. 121, 180505 (2018).


\bibitem{bures}D. Bures, An extension of Kakutani's theorem on infinite product measures to the tensor product of semifinite $ w\sp{\ast}$-algebras, Trans. Amer. Math. Soc. 135, 199 (1969).
\bibitem{uhl}A. Uhlmann, The ``transition probability'' in the state space of a $*$-algebra, Rep.  Math. Phys.
9, 273 (1976).


\bibitem{woot2}W. Wootters, A Measure of the Distinguishability of Quantum States. In: P. Meystre, M. O.  Scully (eds), Quantum Optics, Experimental Gravity, and Measurement Theory. NATO Adv. Sc. Inst. Series 94 (1983).


\bibitem{unc}P. Busch, On the energy-time uncertainty relation. Part I: Dynamical time and time indeterminacy, 
Found. of Phys. 20, 1 (1990).
\bibitem{speed}S. Deffner and S. Campbell, 
Quantum speed limits: from Heisenberg's uncertainty principle to optimal quantum control, J. Phys. A: Math. Theor. 50, 453001 (2017).

\bibitem{luo}S. Luo, Wigner-Yanase Skew Information and Uncertainty Relations, Phys. Rev. Lett. 91, 180403 (2003).

\bibitem{gibi}P. Gibilisco and T. Isola, On a refinement of Heisenberg uncertainty relation by means of quantum Fisher information, J. Math. Anal. App. 375, 270 (2011).
\bibitem{speed2}
D. Paiva Pires, M. Cianciaruso, L. C. Celeri, G. Adesso, and D. O. Soares-Pinto, Phys. Rev. X 6, 021031 (2016).
\bibitem{speed3}C. Zhang, B. Yadin, Z.-B. Hou, H. Cao, B.-H. Liu, Y.-F. Huang, R. Maity, V. Vedral, C.-F. Li, G.-C. Guo, and D. Girolami, Phys. Rev. A 96, 042327 (2017).

\bibitem{monta}A. Montanaro, Quantum algorithms: an overview, npj Quantum Information 2, 15023 (2016). 
\bibitem{tomo} K. Banaszek, M. Cramer,  and D. Gross, Focus on Quantum Tomography, New J. Phys. 15, 125020 (2013).

\bibitem{notepermutations}Different reshufflings of the same setup define different partitions. Given a three-partite system $ABC$, we can implement three partitions of the $\left\{\tilde{{\cal M}}^2,\tilde{{\cal M}}^1\right\}$ type:  $\left\{\tilde{{\cal M}}^{AB}, \tilde{{\cal M}}^C\right\}, \left\{\tilde{{\cal M}}^{AC}, \tilde{{\cal M}}^B\right\}, \left\{\tilde{{\cal M}}^{BC}, \tilde{{\cal M}}^A\right\}$.

 \bibitem{epaps}Supplementary Material.
\bibitem{notebures}%The normalization factor $2/\pi$ makes the the distance to take  maximal value one. 
Note that the Bures length is different from the ``Bures distance'', which is defined as $2-2\,F(\rho_N,\sigma_N)$. 


 \bibitem{brau}S. L. Braunstein and C. M. Caves, Statistical distance and the geometry of quantum states, Phys. Rev. Lett. 72, 3439 (1994).
%\bibitem{fuchs}C. A. Fuchs and J. van de Graaf, Cryptographic distinguishability measures for quantum-mechanical states, IEEE Trans. Inf. Theory 5, 1216 (1999).
%\bibitem{geo2}H. Araki, A remark on Bures distance function for normal states, Publ. RIMS Kyoto Univ. 6, 477(1970).

\bibitem{flammia}S. T. Flammia and Y.-K. Liu, Direct fidelity estimation from few pauli measurements, Phys. Rev. Lett. 106,
230501 (2011).

\bibitem{overlap}L. Cincio, Y. Subaşi, A.a T. Sornborger, and P. J. Coles, Learning the quantum algorithm for state overlap, New J. Phys. 20, 113022 (2018).


\bibitem{fido}H.-Y. Huang, R. Kueng, and J. Preskill, Predicting many properties of a quantum system from very few measurements, Nature Physics 16, 1050 (2020).
 
 

 
  \bibitem{dicke}J. K. Stockton, J. M. Geremia, A. C. Doherty, and H. Mabuchi, Characterizing the entanglement of symmetric many-particle spin-$\frac12$ systems,
Phys. Rev. A 67, 022112 (2003).



\bibitem{metrology}V. Giovannetti, S. Lloyd, and  L. Maccone, Advances in quantum metrology, Nature Phot. 5, 222 (2011).


\bibitem{open}H.-P. Breuer and F. Petruccione, The Theory of Open Quantum Systems, Oxford University Press (2007).


\bibitem{amari}S. Amari and H. Nagaoka, Methods of Information Geometry, American Mathematical  Society (2007).

\bibitem{petz}D. Petz and C. Ghinea, Introduction to Quantum Fisher Information, Quantum Probability and White Noise Analysis, Quantum Probability and Related Topics 27, 261 (2011).



\bibitem{noteiso}Calling $m_i$ the multiplicities of the output state eigenvalues, there are $2^N!/\Pi_i m_i!$ potential $\tau_N$, which can be transformed into each other by eigenvalue permutations. We assume to pick the closest one to the output state.
%%%%%%%%%%,
\bibitem{me}D. Girolami, How Difficult is it to Prepare a Quantum State?, Phys. Rev. Lett. 122, 010505 (2019).

\bibitem{toth}G Tóth and I Apellaniz, Quantum metrology from a quantum information science perspective, J. Phys. A: Math. and Th. 47, 424006 (2014).



\bibitem{semi}S. Boixo, S. T. Flammia, C. M. Caves, and J. M. Geremia, Generalized Limits for Single-Parameter Quantum Estimation, Phys. Rev. Lett. 98, 090401 (2007).

  % \bibitem{note}For example, a bilocal measurement ${\cal M}^1\otimes {\cal M}^1$ is a bipartite measurement ${\cal M}^2$.
   
 
\bibitem{aar}S. Aaronson, Multilinear formulas and skepticism of quantum computing,  Proc. 36th ann. ACM symp. on Th. of comp. 15, 118 (2004).
 \bibitem{nielsengeo}M. A. Nielsen, M. R. Dowling, M. Gu, and A. C. Doherty, Quantum Computation as Geometry, Science 311, 1133 (2006).
 \bibitem{susk}A. R. Brown and L. Susskind, The Second Law of Quantum Complexity, Phys. Rev. D 97, 086015 (2018).
 
 \bibitem{vol}N. Moll {\it et al.}, Quantum optimization using variational algorithms on near-term quantum devices, Quantum Sci. Technol. 3,  030503 (2018).
 
 
 \bibitem{delgado}O. Viyuela, A. Rivas, and M. A. Martin-Delgado, Uhlmann Phase as a Topological Measure for One-Dimensional Fermion Systems, Phys. Rev. Lett. 112, 130401 (2014).
 
%  \bibitem{vn}J. von Neumann, {\it Mathematical Foundations of Quantum Mechanics}, Princeton University Press (1955).

   
\end{thebibliography}
\end{document}